\documentclass[aps,prb,twocolumn,superscriptaddress,showpacs,letter]{revtex4}

\usepackage{graphicx}

\newcommand{\ba}{BaFe$_2$As$_2$}
\newcommand{\sr}{SrFe$_2$As$_2$}
\newcommand{\ef}{E$_F$}
\newcommand{\kf}{k$_F$}
\newcommand{\kz}{k$_z$}
\newcommand{\tsdw}{T$_{\tiny{\textrm{SDW}}}$}
\newcommand{\g}{$\Gamma$}
\newcommand{\mub}{$\mu_{\tiny{\textrm{B}}}$}

\begin{document}

\title{Unconventional electronic reconstruction in undoped (Ba,Sr)Fe$_2$As$_2$ across the spin density wave transition}

\author{M. Yi}
\affiliation{Stanford Institute for Materials and Energy Sciences, SLAC National Accelerator Laboratory, 2575 Sand Hill Road, Menlo Park, CA 94025, USA}
\affiliation{Geballe Laboratory for Advanced Materials, Departments of Physics and Applied Physics, Stanford University, CA 94305, USA}
\author{D. H. Lu}
\affiliation{Stanford Synchrotron Radiation Lightsource, SLAC National Accelerator Laboratory, 2575 Sand Hill Road, Menlo Park, CA 94025, USA}
\author{J. G. Analytis}
\author{J.-H. Chu}
\affiliation{Stanford Institute for Materials and Energy Sciences, SLAC National Accelerator Laboratory, 2575 Sand Hill Road, Menlo Park, CA 94025, USA}
\affiliation{Geballe Laboratory for Advanced Materials, Departments of Physics and Applied Physics, Stanford University, CA 94305, USA}
\author{S.-K. Mo}
\affiliation{Advanced Light Source, Lawrence Berkeley National Laboratory, Berkeley, CA 94720, USA}
\author{R.-H. He}
\affiliation{Stanford Institute for Materials and Energy Sciences, SLAC National Accelerator Laboratory, 2575 Sand Hill Road, Menlo Park, CA 94025, USA}
\affiliation{Geballe Laboratory for Advanced Materials, Departments of Physics and Applied Physics, Stanford University, CA 94305, USA}
\author{M. Hashimoto}
\affiliation{Advanced Light Source, Lawrence Berkeley National Laboratory, Berkeley, CA 94720, USA}
\author{R. G. Moore}
\affiliation{Stanford Synchrotron Radiation Lightsource, SLAC National Accelerator Laboratory, 2575 Sand Hill Road, Menlo Park, CA 94025, USA}
\author{I. I. Mazin}
\affiliation{Center for Computational Materials Science, Naval Research Laboratory Laboratory, Washington, District of Columbia 20375, USA}
\author{D. J. Singh}
\affiliation{Materials Science and Technology Division, Oak Ridge National Laboratory, Oak Ridge, TN 37831-6114, USA}
\author{Z. Hussain}
\affiliation{Advanced Light Source, Lawrence Berkeley National Laboratory, Berkeley, CA 94720, USA}
\author{I. R. Fisher}
\affiliation{Stanford Institute for Materials and Energy Sciences, SLAC National Accelerator Laboratory, 2575 Sand Hill Road, Menlo Park, CA 94025, USA}
\affiliation{Geballe Laboratory for Advanced Materials, Departments of Physics and Applied Physics, Stanford University, CA 94305, USA}
\author{Z.-X. Shen}
\email{zxshen@stanford.edu}
\affiliation{Stanford Institute for Materials and Energy Sciences, SLAC National Accelerator Laboratory, 2575 Sand Hill Road, Menlo Park, CA 94025, USA}
\affiliation{Geballe Laboratory for Advanced Materials, Departments of Physics and Applied Physics, Stanford University, CA 94305, USA}

\date{\today}

\begin{abstract}
Through a systematic high resolution angle-resolved photoemission study of the iron pnictide compounds (Ba,Sr)Fe$_2$As$_2$, we show that the electronic structures of these compounds are significantly reconstructed across the spin density wave ordering, which cannot be described by a simple folding scenario of conventional density wave ordering. Moreover, we find that LDA calculations with an incorporated suppressed magnetic moment of 0.5\mub~can match well the details in the reconstructed electronic structure, suggesting that the nature of magnetism in the pnictides is more itinerant than local, while the origin of suppressed magnetic moment remains an important issue for future investigations.
\end{abstract}


\maketitle

\section{Introduction}
The iron pnictides are a promising class of high temperature superconductors after the cuprates. One striking similarity between the two is that both have an antiferromagnetic ground state for the parent compounds, and that superconductivity emerges as the magnetic ordering is suppressed, with doping in cuprates, and with doping or other means in pnictides. The fact that both classes of high temperature superconductors exist in proximity to a magnetic ground state renders the understanding of this magnetic ground state a crucial matter towards the ultimate understanding of the origin of high temperature superconductivity. No surprise that from the very beginning two key issues have been hotly debated in the field: the strength of the electronic correlations and the origin of the magnetism. Unlike the cuprates, whose parent compounds are antiferromagnetic Mott insulators well-described by strong Coulomb correlations with localized magnetic moments and nearest-neighbors superexchange interaction, the iron pnictide parent compound is a bad metal exhibiting spin density wave (SDW) type ordering\cite{delacruz,huang}, and the role and nature of electronic correlations in iron pnictides appears to be very different from that in the cuprates. Early DMFT calculations following the discovery of superconductivity in the pnictides have made contradictory predictions regarding the strength of the correlations and possible proximity to a Mott insulator\cite{haule1,anisimov}. Subsequent DMFT calculations\cite{anis-review,aichhorn} as well as spectroscopic measurements\cite{anis-review,tom} showed that iron pnictides are in weak to moderate correlations regime, and that the original confusion was due to an incomplete account of the Fe-As hybridization\cite{anis-review,aichhorn} and the emphasis on the role of the Hund's $J$ (as opposed to the Hubbard $U$). However, there is still no consensus regarding the nature of the magnetism in the parent compounds of iron pnictides, whether it is of local nature\cite{dai,fang,xu,ma} or is dominated by itinerant character\cite{dong,yildirimC,mazinNatPhys,kariyado,han,yaresko}. Recently it was argued that the moments form in a local manner, driven by strong Hund's rule coupling in Fe, but their interaction is of one-electron origin, more of an itinerant nature\cite{mazin2}. A comprehensive understanding of magnetic effects in parent compounds serves as a good starting point towards ultimate understanding of the physics of superconductivity in iron pnictides.

Angle-resolved photoemission spectroscopy (ARPES) can play an important role in advancing our understanding of these two issues by providing detailed information of the electronic structures. Recent ARPES measurement on the high temperature paramagnetic state of parent compound \ba~has revealed a band structure that matches reasonably well with the local-density approximation (LDA) calculations after a momentum-dependent shift and renormalization\cite{yi}. In contrast, limited ARPES studies on the low temperature SDW ordering state of the parent compounds\cite{feng,feng2,hasan,zhou,shin,kaminski} have been less conclusive, leading to different interpretations of the complex electronic structures in the SDW state. It was pointed out that some spectra resemble ferromagnetically exchange split bands\cite{feng}, even though the magnetic order in the bulk is antiferromagnetic and it is highly unlikely that ferromagnetism may exist on the surface in these materials. Others came up with a  more expected folding-nesting picture\cite{hasan}. However, the antiferromagnetic exchange gaps found in that work were substantially smaller then what one would have expected given a 1\mub~magnetic moment on Fe, and some bands seem to see no effect of the SDW ordering at all, which is hard to reconcile with a well established 90\% loss of the Drude weight in optics\cite{pfuner,nlwang} and a large loss of carriers in the Hall experiments\cite{Hall}. Below we present a systematic high resolution ARPES study of the SDW ordering in the pnictide parent compounds (Ba,Sr)Fe$_2$As$_2$, with unprecedented details that lead us to different understanding of the complex electronic structures in the SDW state. By comparing the measured band dispersion to LDA calculations incorporating different fixed magnetic moments, we show that the electronic structure in the SDW state is comparable to LDA with a magnetic moment of 0.5\mub. The ability of LDA with fixed magnetic moment to match details in the electronic structures of the iron pnictides even in the SDW state suggests that itinerant physics is the more appropriate starting point for the pnictides, albeit with finite correlation physics as suggested by a momentum-dependent shift and renormalization of the LDA needed to obtain such a match. Furthermore, the origin of the much suppressed magnetic moment is likely due to the spin fluctuations that are not taken into account in LDA calculations. A comprehensive theory beyond mean-field approach is needed to fully describe the magnetic ground state of the parent compound.

\section{Methods}
\subsection{Sample Preparation and Measurement}
High quality single crystals of (Ba,Sr)Fe$_2$As$_2$ were grown using the flux method\cite{chu}. ARPES measurements were carried out at both beamline 5-4 of the Stanford Synchrotron Radiation Lightsource and beamline 10.0.1 of the Advanced Light Source using SCIENTA R4000 electron analyzers. The total energy resolution was set to 13 meV or better and the angular resolution was 0.3$^{\circ}$. Single crystals were cleaved in situ at 10 K for all low temperature and temperature dependence measurements and at 150 K for high temperature measurements. ARPES spectra were obtained using 25 eV photons unless otherwise noted. All measurements were done in ultra high vacuum chambers with a base pressure better than 4x10$^{-11}$ torr.

\subsection{\label{sec:secLDA}Electronic Structure Calculation}
All calculations were performed using the full potential linearized augmented plane-wave (LAPW) method as implemented in the WIEN2k package\cite{wien2k}. The experimental crystal structure for \ba~was used for all calculations. The same setup was used as in Ref. 14, except that the local density approximation without gradient corrections was used for all calculations. In order to suppress the magnetic moment to a desired value, we used LDA+U technique with J=0 and U$<$0. We stress that there is no physical meaning in this procedure and no physical justification. It is simply a way to artificially reduce the tendency to magnetism within a mean-field approach and mimic the effect of spin fluctuations. We verified that by increasing $\mid$U$\mid$ we can drive the system to a practically non-magnetic state, whose band structure is nearly identical to that obtained from nonmagnetic calculations without U. This gives us confidence that no discernible side effects from the artificial U term, besides suppression of the Fe moment, appear in the calculations.

\section{Electronic Structure Measurements}
\begin{figure}[t]
\includegraphics[width=0.48\textwidth]{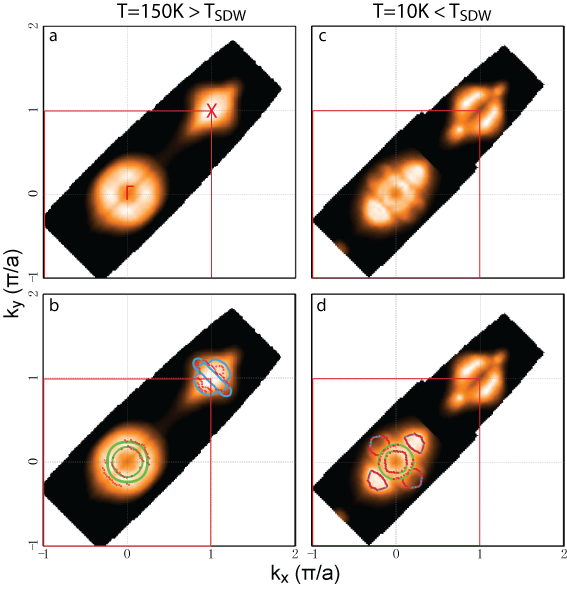}
\caption{\label{fig:fig1}Comparison of measured FS of \ba~above and below \tsdw. (a) FS intensity map taken above \tsdw~(T = 150 K) with an integration window of 5 meV about \ef. (b) FS map from (a) overlaid with \kf~points (points in momentum where bands cross \ef) in red and FS outlines based on LDA calculations. (c)-(d) same as those in (a)-(b) except taken below \tsdw~(T = 10 K). Green outlines denote hole pockets while blue outlines denote electron pockets. $\Gamma$ and X points refer to notation in the BZ corresponding to the 2-Fe unit cell. The red squares mark the first BZ in the paramagnetic state.}
\end{figure}

\begin{figure*}[tbh]
\includegraphics[width=0.97\textwidth]{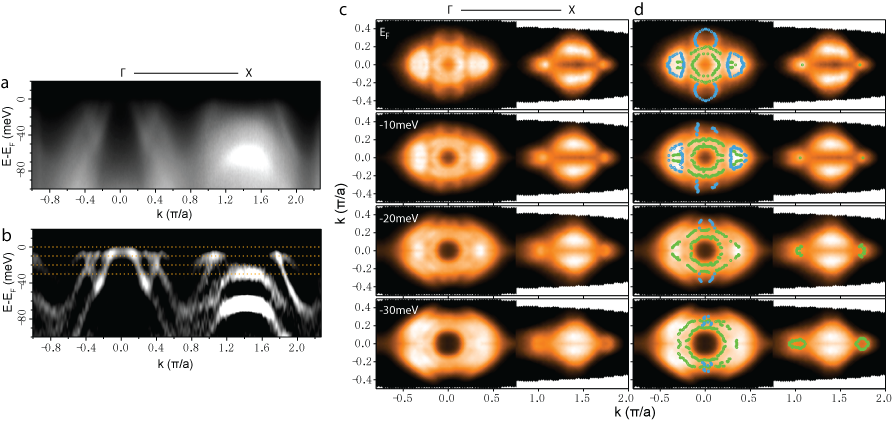}
\caption{\label{fig:fig2}Constant energy mapping of \ba~in the SDW state. (a) Raw ARPES spectra of \ba~along the \g-X high symmetry cut. (b) Second derivative image of the spectra in (a) for better contrast of the measured band dispersions. The dotted lines mark the binding energy levels for the constant energy mappings in (c)-(d). (c) Constant energy maps across the \g-X region at E$_B$ = 0, 10meV, 20meV, and 30meV. (d) Maps in (c) overlaid with dots marking the constant energy contours of the bands. Green marks denote hole-like features while blue marks denote electron-like features.}
\end{figure*}

The parent compound \ba~(\sr) undergoes both an SDW ordering and tetragonal to orthorhombic structural transition when cooled across \tsdw~= 137 K (200 K) marked by an anomaly in transport measurements\cite{rotter,ni,chu,yan} and confirmed by neutron measurements\cite{huang,zhao,kaneko}. Figure~\ref{fig:fig1} contrasts the measured Fermi surfaces (FS) of \ba~above and below \tsdw. In the paramagnetic state (Fig.~\ref{fig:fig1}(a)-(b)), there are three hole pockets centered around the \g~point, including two nearly degenerate as predicted by LDA but unresolvable in the ARPES data. At the X point (($\pi$, $\pi$) in the Brillouin zone (BZ) corresponding to the 2-Fe unit cell), two electron pockets hybridize to form the FS\cite{yi}. Below \tsdw, the FS is drastically reconstructed. As shown in Fig.~\ref{fig:fig1}(c)-(d), there are two small pockets centered at \g, surrounded by four petal-like pockets.

The character of FS pockets as measured using ARPES can usually be easily determined by tracing the dispersion of associated bands. In a multi-band system with a complex FS topology, it is more practical to examine the constant energy mapping as electron (hole) FS pockets shrinks (expands) in area towards higher binding energy. In Fig.~\ref{fig:fig2}, we show a series of constant energy maps of \ba~in the SDW state across both \g~and X regions at increasing binding energies from the Fermi level (\ef). Figure~\ref{fig:fig2}(b) is the second derivative of the raw ARPES spectra (Fig.~\ref{fig:fig2}(a)) taken along the \g-X high symmetry cut, showing the band dispersions and the binding energy levels selected for the four mappings in Figs.~\ref{fig:fig2}(c)-(d). In Fig.~\ref{fig:fig2}(d), we plot on top of the mappings points where bands cross the corresponding binding energies, outlining the evolution of the constant energy contour as a function of binding energy. Around the \g~point, the two inner features (green) grow towards higher binding energy, and are therefore hole-like. The four surrounding petals shrink towards higher binding energy and quickly disappear, hence electron-like. Note also another hole-like feature appearing on the outside of the petals below \ef. At the X point, two small hole-like features at higher binding energies shrink towards point-like bright spot features at \ef~along the \g-X high symmetry line, while the shape of the two lobes of intensity in between are harder to determine using this method alone, and will be discussed later with the help of the corresponding band dispersions. The Luttinger volumes of the outer and inner hole pockets at \g~are 3.4$\%$ and 1.4$\%$ of the BZ, respectively, and that of the electron-like petals are 0.8$\%$. The last two values are in good agreement with quantum oscillation measurements~\cite{analytis,sebastian} in which FS pockets with sizes 1.7$\%$ and 0.7$\%$ were reported. Note that the pocket (3.4$\%$) without a corresponding size in quantum oscillations appears to be open approaching the \g-X symmetry line as indicated by those \kf~points marked in red in Fig.~\ref{fig:fig1}(d)). This could be due to either orbital symmetry-related spectral weight suppression resulting from the experimental geometry, or the opening of a gap on this portion of the FS, in which case the pocket is not closed and hence undetectable in quantum oscillations. In addition, quantum oscillations also observed a smaller 0.3$\%$ pocket, which could correspond to the lobe-like FS near the X point, whose exact shape and size is harder to resolve in the ARPES data.

\begin{figure*}[t]
\includegraphics[width=0.97\textwidth]{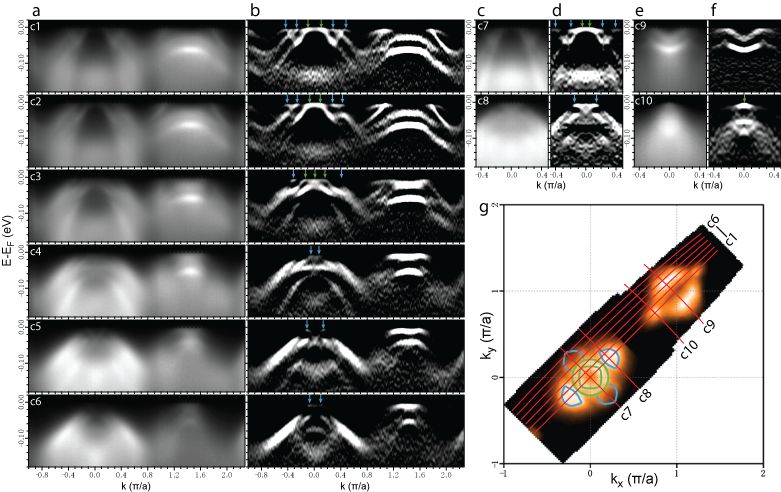}
\caption{\label{fig:fig3}Momentum-dependent band dispersion of \ba~in the SDW state. (a), (c), (e), Raw spectra images of cuts across the BZ as marked in (g). (b), (d), (f), Second derivative plots of the corresponding spectra to the left showing the band dispersions in better contrast. Arrows at \ef~mark \kf~points of the corresponding FS in (g). (g) FS in the SDW state. All data were taken at 10 K.}
\end{figure*}

\begin{figure}[t]
\includegraphics[width=0.47\textwidth]{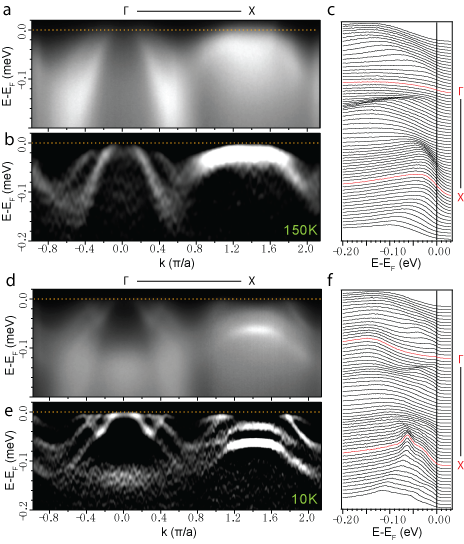}
\caption{\label{fig:fig4a} Contrast of \ba~band dispersions across the \tsdw. (a) Raw spectra image along the \g-X high symmetry cut above \tsdw~(T = 150 K). (b) Second derivative image of (a). (c) Energy Distribution Curves (EDCs) of the cut in (a). (d)-(f) same as in (a)-(c) but taken below \tsdw~(T = 10 K).}
\end{figure}

More information about the reconstructed electronic structure can be obtained from a study of detailed band dispersions across the BZ (Fig.~\ref{fig:fig3}). In Fig.~\ref{fig:fig3}(a)-(f) is a series of cuts parallel or perpendicular to the \g-X high symmetry line. Centered on the \g~point are three hole-like bands, two of which cross \ef~while the third outermost one joins with an electron-like feature at \ef~near \g~and extends towards X. The two \kf~points of the inner hole pocket merge to a single point by c3. The outer hole pocket is suppressed along the high symmetry cut, but becomes apparent in c3. There is also an electron-like dispersion centered at \g, which crosses the hole bands on the high symmetry cut and hybridize with them away from the high symmetry cuts, opening up hybridization gaps where they cross. Note that such an electron-like band is absent around \g~in the high temperature paramagnetic state. The \kf~points of these hybridized dispersions form the petal-like pockets on the FS. At the X point, at around 30 meV and 65 meV below \ef~there are two saddle-like bands with hole-like dispersions in the \g-X direction (c1), and electron-like dispersions in the perpendicular direction (c9). As we go parallel and away from the \g-X high symmetry line (c1-c6), these two bands approach \ef, and the upper band eventually touches \ef, with its flat band top contributing to the two lobes of intensity seen around the X point in the FS map (Fig.~\ref{fig:fig1}(d)), as can be seen in the wing-like dispersion near \ef~in the perpendicular cut c9. In c10, we see a hole-like feature that is responsible for the bright spots along the \g-X direction in the FS map. Compared to the band dispersions in the paramagnetic state, there is significantly more complexity in the SDW state (Fig.~\ref{fig:fig4a}), such as the prominent band splitting at the X point and the electron-like feature around the \g~point that are absent in the paramagnetic state.

\begin{figure}
\includegraphics[width=0.47\textwidth]{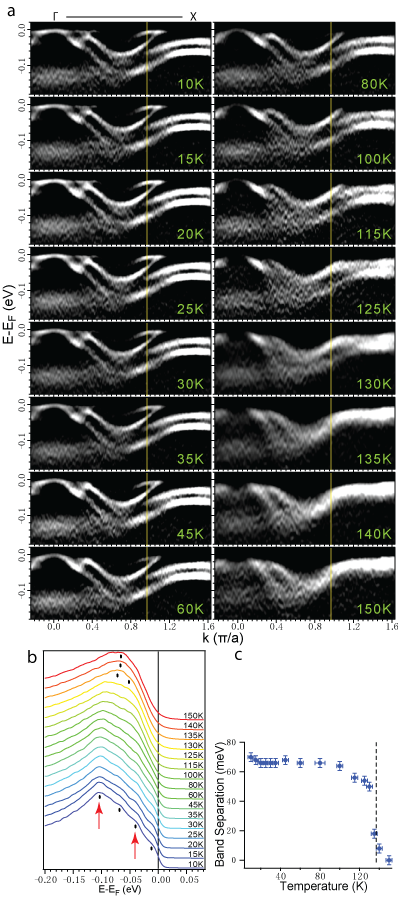}
\caption{\label{fig:fig4}Temperature evolution of \ba~band dispersions across the \tsdw. (a) Second derivatives of spectra along the \g-X high symmetry cut showing the temperature evolution of band dispersions across \tsdw. (b) Temperature dependence of a single EDC (marked by the yellow lines in (a)) across \tsdw. The peak positions mark the band binding energy positions at this momentum. (c) Separation of the two bands in energy (marked by arrows in (b)) as a function of temperature across \tsdw~= 137 K.}
\end{figure}

\begin{figure*}[tbh]
\includegraphics[width=0.95\textwidth]{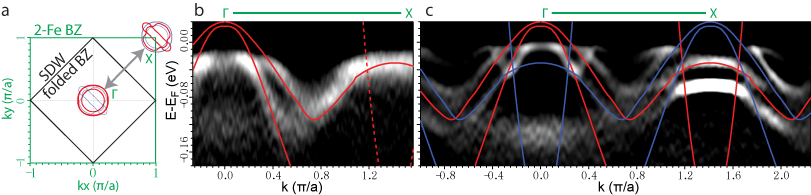}
\caption{\label{fig:fig5a}Simple Folding Scenario. (a) Simple folding schematic. Direct ($\pi$, $\pi$) folding of the paramagnetic BZ (green) into the SDW BZ (black) means the folded FS (blue) on top of the paramagnetic FS (red). (b) Second derivative plot of \ba~band dispersions measured in the paramagnetic state (T = 150 K) along \g-X with eye guides. (c) \ba~band dispersions measured in the SDW state (T = 10K) along \g-X. Red lines denote high temperature band dispersion from (b). Blue lines are direct ($\pi$, $\pi$) folding of the high temperature bands. The simple folding scheme does not reproduce measured dispersions.}
\end{figure*}

\begin{figure*}[tbh]
\includegraphics[width=0.95\textwidth]{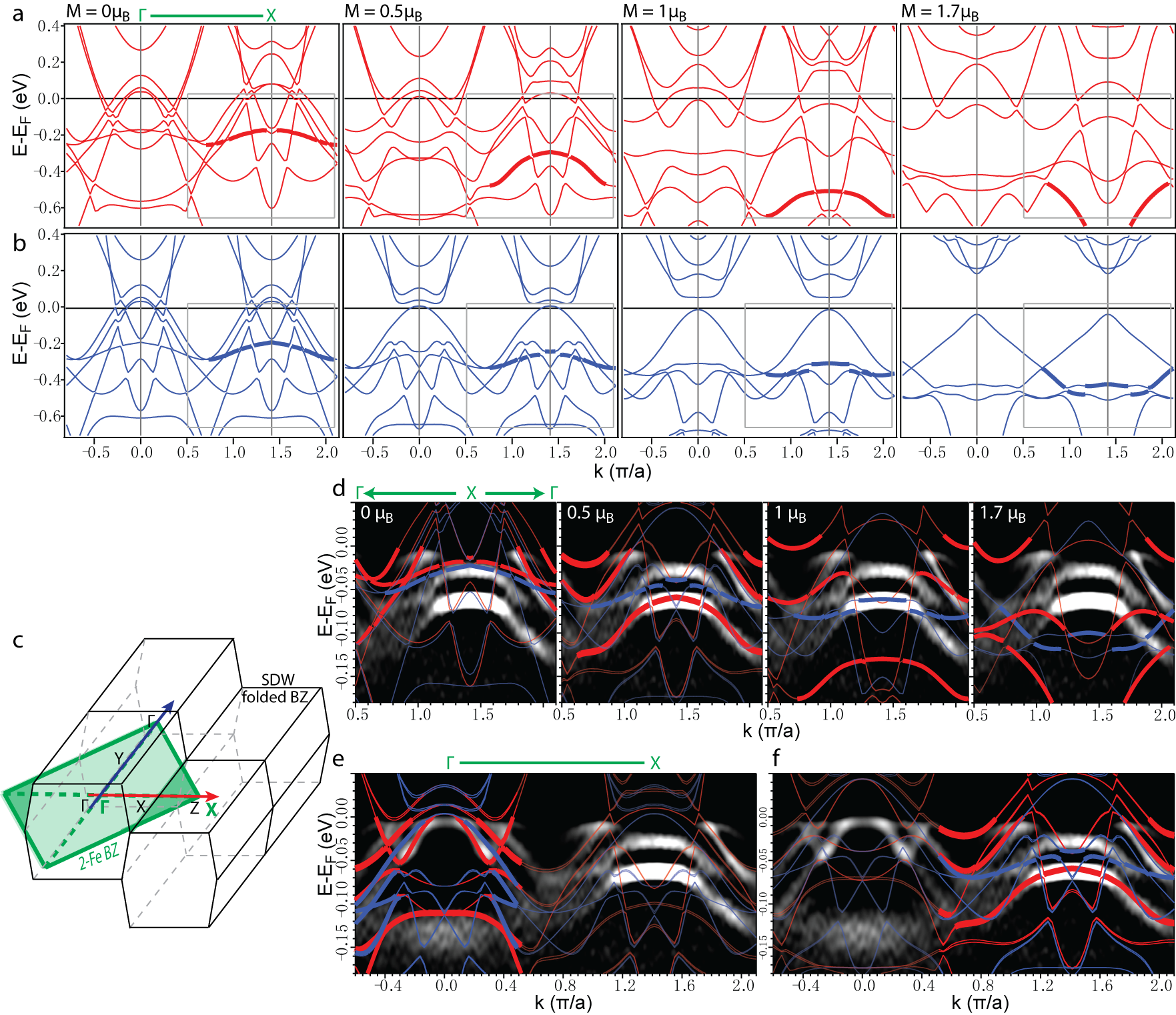}
\caption{\label{fig:fig5b}Comparison of \ba~band dispersions with fixed-moment LDA. (a) LDA calculations of bands along \g-X-Z in the folded SDW BZ (corresponding to \g-X of the paramagnetic BZ, see (c)) with magnetic moments fixed at different values. Thicker segments trace the trend that bands shift in energy with increasing moment. (b) Same as in (a) except for the direction along \g-Y-\g~in the folded SDW BZ. (c) 3-D BZ schematic. Green denotes 1st BZ in the paramagnetic state that is probed under our experimental setup. In the SDW state, the base-centered orthorhombic structure corresponds to a BZ (black) staggered in \kz~in one direction. The original equivalent \g-X directions in the paramagnetic state now becomes \g-X-Z (red arrow) and \g-Y-\g~(blue arrow) directions. Averaging over magnetic domains under the synchrotron beam spot leads to a mixture of signals from the \g-X-Z and \g-Y-\g~directions. (d) LDA (\kz~=0) with various fixed magnetic moments overlaid on experimental data around X point in the \g-X direction, with contribution from both \g-X-Z (red) and \g-Y-\g~(blue) directions. All LDA bands are shifted up in energy by 0.12 eV and renormalized by a factor of 3. (e) The best match (LDA with M = 0.5\mub) overlaid on band dispersion along the \g-X cut, unshifted but renormalized by a factor of 3 to fit with bands around \g. (f) same as (e) except with LDA shifted up in energy by 0.12 eV and renormalized by the same factor of 3 to fit with bands around X.}
\end{figure*}

To see how such reconstructed bands evolve into the much simpler bands in the normal state, we present a detailed temperature-dependent study across \tsdw~(Fig.~\ref{fig:fig4}). At low temperatures, at least four separate bands can be clearly resolved around the X point, which merge together as soon as \tsdw~is crossed (Fig.~\ref{fig:fig4}(a)). Fig.~\ref{fig:fig4}(b) traces a representative energy distribution curve (EDC) at a fixed momentum as a function of temperature. A close look at the EDC's taken right above and below \tsdw~(140 K vs. 135 K) highlights the abrupt change in the band dispersions across the SDW transition, which cannot be accounted for by a trivial thermal broadening or a broadening due to a simple change in scattering rate. To better visualize the evolution of the electronic structure across the SDW transition, we plot in Fig.~\ref{fig:fig4}(c) the energy separation of the two most prominent bands around X as a function of temperature, which shows a temperature dependence reminiscent of the ordered magnetic moment measured by neutron experiments\cite{huang,zhao,kaneko}.

\section{Discussion}
\subsection{Simple Folding Scenario}
To understand the dramatic change in the electronic structure across \tsdw, we begin by considering a simple folding scenario, which was suggested in a previous ARPES report\cite{hasan}. In the low temperature state, \ba~has an orthorhombic structure. Neglecting the small orthorhombic distortion and \kz~dispersion, it has a BZ that is $\sqrt{2}\times\sqrt{2}$ folded from the paramagnetic BZ (Fig. ~\ref{fig:fig5a}(a)), where the \g~and X points are folded onto each other. In a simple folding picture as in the case of CDW in binary rare earth tellurides\cite{brouet} or SDW in chromium\cite{rotenberg}, additional bands in the ordered state are generated by a simple translation of the original bands via \textbf{Q}$_{CDW}$ or \textbf{Q}$_{SDW}$, with hybridization gaps opening where folded bands and original bands cross, resulting in loss of considerable portions of the FS. In Fig.~\ref{fig:fig5a}(b)-(c) we consider the possibility of such a scenario by plotting a direct folding of the measured paramagnetic bands on the measured dispersion in the SDW state, but find that such a direct folding cannot reproduce the measurement with or without an opening of gaps of any size. Apparently, there is significant reconstruction of the electronic structure inexplainable by a simple folding picture, as suggested by first principle calculations\cite{mazin2}.

\begin{figure*}[tbh]
\includegraphics[width=0.97\textwidth]{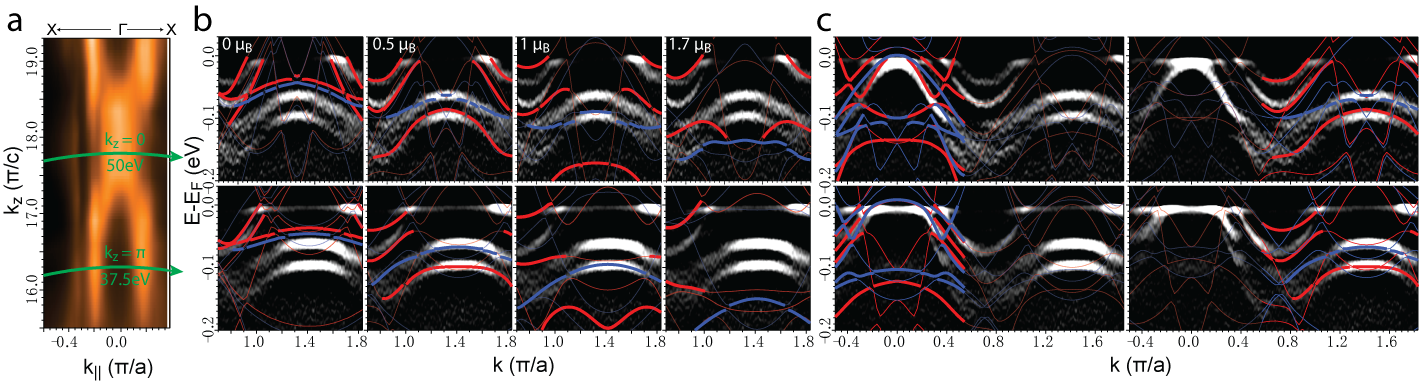}
\caption{\label{fig:fig6}Comparison of \sr~band dispersions with fixed-moment LDA. (a) Measured FS on the \kz-k$_{\parallel}$ (X-\g-X) plane showing \kz~dispersion probed by varying incident photon energy (T = 10 K). (b) LDA with various fixed magnetic moments overlaid on band dispersions around X point in the \g-X direction, with contribution from both \g-X-Z (red) and \g-Y-\g~(blue) directions due to domain mixing. Top panels: data taken with 50 eV photons, corresponding to \kz~= 0 in the BZ, and overlaid with LDA calculations for \kz~= 0. Bottom panels: data taken with 37.5 eV photons, corresponding to \kz~= $\pi$ in the BZ, and overlaid with LDA calculations for \kz~= $\pi$. All LDA bands are shifted up in energy by 0.08 eV and renormalized by a factor of 2.5. (c) the best match (LDA with M = 0.5\mub) overlaid on band dispersion along the \g-X cut. Top row: data taken with 50 eV photons (\kz~= 0) overlaid with LDA (\kz~= 0). Bottom row: data taken with 37.5 eV photons (\kz~= $\pi$) overlaid with LDA (\kz~= $\pi$). Left panels: LDA shifted down in energy by 0.01 eV and renormalized by 2.5 to match with bands around \g. Right panels: LDA shifted up in energy by 0.08 eV and renormalized by the same factor to match with bands around X.}
\end{figure*}

\subsection{Comparison with LDA}
Next, we turn our attention to LDA, which has been quite successful in capturing details in the electronic structures of both nonmagnetic LaFePO\cite{lu} and \ba~in the paramagnetic state\cite{yi}. To extend the nonmagnetic LDA calculations for the paramagnetic states to the SDW state, one has to correctly take into account the effect of the ordered magnetic moment in the SDW state. In this regard, LDA has been known to overestimate\cite{mazin,yin,yildirim,yaresko,mazinNatPhys} the magnetic moments in the iron arsenides compared with experimental probes such as neutron. Hence, we use a negative U method as described in Sec.~\ref{sec:secLDA} to artificially stabilize solutions for various suppressed moments in order to compare with our measurements. In Fig.~\ref{fig:fig5b}(a)-(b) we show the results of such LDA calculated bands for various moments (0 to 1.7\mub) along two high symmetry cuts. The calculated bands are clearly sensitive to the magnetic moment in that they shift in energy with increasing moment. In addition, the difference between the \g-X-Z and \g-Y-\g~directions becomes more dramatic with increasing moment. These two trends are better seen by tracing the thickened segments.

We then compare the LDA calculated bands to our measured dispersions. We note that in the SDW state, the k$_x$ and k$_y$ directions in the SDW BZ are no longer equivalent due to the orthorhombic distortion and the collinear antiferromagnetic order, i.e., the two orthogonal \g-X directions in the paramagnetic BZ would correspond to \g-X-Z and \g-Y-\g~directions in the SDW BZ (Fig.~\ref{fig:fig5b}(c)). On the other hand, due to the existence of twin domains\cite{canfield}, ARPES spectra taken along the \g-X direction in the paramagnetic BZ would correspond to a mixture of signals from both \g-X-Z (Fig.~\ref{fig:fig5b}(a)) and \g-Y-\g~(Fig.~\ref{fig:fig5b}(b)) directions as the beam spot is expected to cover multiple domains. Therefore, we still adopt the notation of the paramagnetic BZ and plot LDA calculated bands from both \g-X-Z and \g-Y-\g~directions on top of the corresponding ARPES spectra around X (Fig.~\ref{fig:fig5b}(d)). With an upward shift of 0.12 eV and a bandwidth renormalization factor of 3, the LDA calculations with M = 0.5\mub~comes to a close match with the measured band dispersions as shown by the highlighted bands. Note that the LDA bands not seen in the ARPES spectra are the electron bands and lower hole band corresponding to those suppressed under this polarization geometry due to orbital symmetry, as reported previously for the paramagnetic state\cite{yi,fengorb}. The LDA bands for the other three moments have been shifted and renormalized by the same amount and the corresponding bands are highlighted for comparison. The different curvature and relative band positions for the other three moments cannot be made to match the measured band dispersion with any combination of shift and renormalization as with the way for M = 0.5\mub. Moreover, in Fig.~\ref{fig:fig5b}(e), we extent the comparison of LDA with M = 0.5\mub~to the \g~point as well, for which a match is obtained with no shift and the same renormalization factor of 3.

Furthermore, this match is robust to \kz~effects as demonstrated on the \sr~system (Fig.~\ref{fig:fig6}). The parent compound \sr~is similar to \ba~in both electronic structure\cite{feng,hasan} and transport properties\cite{yan}, with a slightly higher magnetic moment (M = 0.94\mub) and \tsdw~(200 K). Through a careful photon energy dependence study, we mapped out the \kz~dispersions of \sr~(Fig.~\ref{fig:fig6}(a)), from which we identified two photon energies (50 eV and 37.5 eV) that correspond to \kz~= 0 and $\pi$ at \g, respectively, and compared with LDA of various moments for the corresponding \kz~values (Fig.~\ref{fig:fig6}(b)). Again, the LDA with M = 0.5\mub~is the best match for both photon energies. We then extend the match to \g, where a strong \kz~effect predicted by the calculations is clearly reproduced in the experimental dispersions (Fig.~\ref{fig:fig6}(c)). The degree of the match across \kz~after the same momentum-dependent shift and renormalization of LDA bands is a strong testament of the robustness of the match.

Such a comparison also allows us to gain deeper understanding of the dramatic change of the band dispersions and Fermi surface reconstructions across the SDW transition. One of the reasons for the seemingly complex band dispersions and Fermi surfaces, such as the increased number of bands, is due to the mixture of the signal from two domains. Due to the presence of the collinear antiferromagnetic ordering in the SDW state, the band dispersions along \g-X-Z and \g-Y-\g~directions become quite different, resulting in a pair of separated bands upon domain mixing, such as the highlighted bands in Fig.~\ref{fig:fig5b}(a) and~\ref{fig:fig5b}(b). The separation between such a pair of bands is indicative of the magnitude of the ordered moment, which naturally explains the temperature-dependent ``band splitting'' shown in Fig.~\ref{fig:fig4}. This also explains why a simple band folding is insufficient to describe the drastic changes in electronic structure across the SDW transition: as the ordered magnetic moment develops below the SDW transition, all bands shift substantially.

As evident from the result of this analysis consistent with neutron measurements, LDA calculations substantially (by at least 50\%) overestimate the Fe magnetic moment in the SDW phase. The reason is that LDA is by construction a mean field theory, while in the actual materials magnetic ordering is disrupted by spin fluctuations. These fluctuations are largely transverse, that is, the fluctuating variable is the direction of the magnetic moment\cite{mazinNatPhys,yildirimC}. The band structure reflects the magnetic moment that is averaged over the length scale of several ($\sim 10$) lattice parameters, which is smaller than the local moment. Assuming that the main effect on the band structure is this reduction (a reasonable assumption, albeit still an assumption), we can emulate it by suppressing the magnetic moment with an artificial negative Hubbard U. This procedure simply uses the fact that in LDA+U, the effective Hund's rule coupling $J$ defined as $2 d^2E/dM^2$ is renormalized as $\Delta J=U/5$ (for d electrons)\cite{mazin03}.

While the presented calculations match the experimental data in great details, a remaining issue is the origin of the required momentum-dependent shift. We note here that a very similar momentum-dependent shift and renormalization across the \g~and X points gives a good match between nonmagnetic LDA and band dispersion of \ba~in the paramagnetic state\cite{yi}. The consistency reinforces the possibility of some missing physics, of which we discuss a few here. First of all, LDA is known to overestimate the spread of orbitals and to overestimate hybridization effects. This can, in principle, lead to nonuniform shifts. In MgB$_2$, for instance, to match the experiment exactly one needs to shift $\sigma$ bands with respect to the $\pi$ bands as calculated by LDA by about 160 meV\cite{mazinMgB2}, even though there are no correlation effects in MgB$_2$. A second possible cause is the lack of natural cleavage planes in the ternary 122 pnictides. The cleave for the 122 family, as confirmed by various STM studies\cite{eric,yinyi,plummer,vydia}, most likely occurs in the Ba(Sr) layers, leaving disordered or reordered Ba(Sr) atoms atop a reconstructed As plane, which could possibly result in surface relaxation causing a non-uniform shift of bands as in the case of Sr$_2$RuO$_4$\cite{kyle}. On the other hand, it could also suggest intrinsic correlation physics not accounted for in the LDA calculations. In a recent report by Aichhorn \emph{et al.}\cite{aichhorn}, LDA+DMFT calculations predicted weak correlations with an average mass renormalization of 1.6 and orbital and momentum dependent shift of bands compared to LDA calculations as a result of correlations, in qualitative agreement with our observations. A fourth possibility is the combination of a dominance of interband coupling over intraband coupling and strong particle-hole asymmetry of electronic bands in this multiband system, resulting in relative shifts of the hole and electron bands as discussed by Ortenzi \emph{et al}\cite{ortenzi}. At this point, we leave this as an open question to be addressed for systematic future investigations.

Nonetheless, the degree of agreement suggests that LDA incorporating magnetic moment is a good starting point to describe the pnictides even in the magnetic state, and that the physics is likely more itinerant than local, in contrast to the cuprates. Moreover, the measured band dispersions of \ba~(\sr) correspond to a magnetic moment close to 0.5\mub, a value smaller than the bulk value of 0.87\mub~(0.94\mub) measured by neutron\cite{huang,zhao,kaneko}. This could be an indication that the moment on the cleaved surface is suppressed compared to the bulk. On the other hand, the precise temperature dependence tracked by the change in band dispersions across \tsdw~strongly suggests that at the very least, the surface dynamics are bulk-driven.

\section{Conclusion}
In conclusion, we have presented in detail the reconstructed electronic structure of (Ba,Sr)Fe$_2$As$_2$ in the SDW state by mapping out the FS and band structure across the entire BZ, accompanied by a detailed temperature evolution of the band dispersion across \tsdw. Moreover, we have shown that a simple folding picture is not sufficient to describe the significant reconstruction of the electronic structure, but that the magnetic moment in the SDW state and domain mixing have also to be taken into account. We showed this by comparing measured dispersion with LDA calculated with various magnetic moments and obtained a close match with calculations using M = 0.5\mub, an indication that the surface magnetism is smaller than the bulk, albeit bulk-driven. Overall, the ability of LDA calculations to reproduce details in the measured dispersions in the SDW state of (Ba,Sr)Fe$_2$As$_2$, taken in the global picture together with similar systematic matches for LaFePO\cite{lu} and the paramagnetic state of undoped and doped \ba\cite{yi}, is good evidence suggesting that the governing physics in the pnictides is more itinerant than local in nature. Furthermore, the need of magnetic moment suppression in LDA calculations to match ARPES measurements, consistent with neutron measurements, suggests that effects such as spin fluctuations are also important in the physics of the pnictides.

\begin{acknowledgments}
We thank E. Cappelluti, T. P. Devereaux, W.S. Lee, B. Moritz, H. Yao, and Y. Yin for helpful discussions. ARPES experiments were performed at the Stanford Synchrotron Radiation Lightsource and the Advanced Light Source, which are both operated by the Office of Basic Energy Science, U.S. Department of Energy. The Stanford work is supported by DOE Office of Basic Energy Science, Division of Materials Science and Engineering, under contract DE-AC02-76SF00515. Work at ORNL was supported by the DOE, Division of Materials Sciences and Engineering. MY thanks the NSF Graduate Research Fellowship Program for financial support.
\end{acknowledgments}

\end{document}